\documentclass[11pt,onecolumn,amssymb,nofootinbib]{revtex4}
\usepackage{amsmath, amsthm, amscd, amssymb}
\usepackage{bm}
\usepackage{bbm}

\begin{document}

\title{\bf Modeling the flyby anomalies with dark matter scattering}
\author{Stephen L. Adler}
\email{adler@ias.edu} \affiliation{Institute for Advanced Study,
Einstein Drive, Princeton, NJ 08540, USA.}

\begin{abstract}
We continue our exploration of whether the flyby anomalies can be explained by
scattering of spacecraft nucleons from dark matter gravitationally bound to the earth.
We formulate and analyze a simple model in which inelastic and elastic scatterers
populate shells generated by the precession of circular orbits with normals tilted
with respect to the earth's axis.  Good fits to the data published
by Anderson et al. are obtained.

\end{abstract}

\maketitle

\section{Introduction}
In this paper we follow up our earlier investigation \cite{adler1} of the anomalous geocentric frame
orbital energy changes that are observed during earth flybys of various spacecraft, as
reported by Anderson et al. \cite{anderson}.   Some flybys show energy decreases, and
others energy increases, with the largest anomalous velocity changes of order 1 part
in $10^6$.  While the possibility that these anomalies are artifacts of the orbital
fitting method used in \cite{anderson} is still being actively explored, there is also
a chance that they may represent new physics.   In \cite{adler1} we explored the possibility that the flyby anomalies
result from scattering of spacecraft nucleons from dark matter particles in orbit around the earth, with the
observed velocity decreases arising from elastic scattering, and the observed velocity
increases arising from exothermic inelastic scattering, which can impart an energy impulse to a spacecraft nucleon.  Many
constraints on this hypothesis were analyzed in \cite{adler1}, with the conclusion that the dark matter
scenario is not currently ruled out, but requires dark matter to be non-self-annihilating, with the dark
matter scattering cross section on nucleons much larger, and the dark matter mass much lighter, than
usually assumed.

 However, no attempt was made in \cite{adler1} to construct a model for the spatial and velocity
distribution functions for dark matter populations in earth orbit, to see whether it can fit the flyby data reported
in \cite{anderson}.  Formulating such a model is the aim of the present paper. Our basic assumption is to consider
two populations of dark matter particles, one of which scatters on nucleons elastically, and the other of which scatters inelastically, each with a shell-like distribution of  orbits generated by the precession of a
tilted circular orbit around the earth's rotation axis.    The formulas defining
this model are developed in Sec. II, with details of derivations in Appendices, and the results of numerical fits to the flyby data are given in
Sec. III.  We show that good fits to the data are possible, which leaves dark matter scattering as a viable
candidate for explaining the flyby anomalies, pending further investigation of possible artifactual
explanations\footnote{A parameterized post-Newtonian analysis, given in an unpublished memo in the ``Talks+Memos'' section
of the author's home page,  shows that deviations from Einstein gravity within
the framework of metric theories of gravity obeying the equivalence principle cannot give residual accelerations large
enough to explain the flyby anomalies.} of the flyby data,  and further  experiments aimed at directly detecting dark matter and determining its properties.

\section{Formulas defining the model}

\subsection{Velocity change formulas}

We recall from \cite{adler1} formulas for the velocity change when a spacecraft
nucleon of mass $m_1\simeq
1 {\rm GeV}$ and initial velocity $\vec u_1$ scatters from a primary
dark matter particle of mass $m_2$ and initial velocity $\vec u_2$,
into an outgoing nucleon of mass $m_1$ and velocity $\vec v_1$, and
an outgoing secondary dark matter particle of mass $m_2'=m_2-\Delta
m$ and velocity $\vec v_2$ . The inelastic case corresponds to
$m_2'\not= m_2$, while in the elastic case, $m_2'=m_2$ and $\Delta
m=0$. Under the
assumptions, (i) both initial particles are nonrelativistic, so that
$|\vec u_1|<<c, |\vec u_2|<<c$, (ii) the center of mass
scattering amplitude $f(\theta)$ depends only on the auxiliary polar
angle $\theta$ of scattering, and (iii) in the exothermic inelastic case,   $\Delta m/m_2$ and
$m_2'/m_2$ are both of order unity, a
straightforward calculation gives  the outgoing nucleon velocity
change, averaged over scattering angles.
In the elastic scattering case, with $\Delta m=0$, $m_2'=m_2$,
we have
\begin{equation}\label{eq:el}
\langle \delta \vec v_1\rangle = -2\frac{m_2}{m_1+m_2}(\vec
u_1-\vec u_2)\langle \sin^2(\theta/2) \rangle~~~,
\end{equation}
while in the inelastic case a good approximation is
\begin{equation}\label{eq:inel}
\langle \delta \vec v_1\rangle \simeq \frac{\vec u_1-\vec
u_2}{|\vec u_1-\vec u_2|}     \Bigg( \frac{2 \Delta m ~m_2'} {
 m_1 (m_1+m_2')}\Bigg)^{1/2}c \langle \cos \theta \rangle ~~~,
\end{equation}
with $\langle ... \rangle$ denoting the angular average over the center of
mass differential scattering cross section.
Since $\vec u_1$ and $\vec u_2$ are typically of order 10 ${\rm
km} ~{\rm s}^{-1}$, the velocity change in the inelastic case is
significantly larger than that in the elastic case.

\subsection{Change in outgoing spacecraft velocity}

Again as shown in \cite{adler1},
to get the force per unit spacecraft mass resulting from dark
matter scatters, that is, the acceleration, one multiplies the
velocity change in a single scatter $\langle \delta \vec
v_1\rangle $ by the number of scatters per unit time.  This latter
is given by the flux $|\vec u_1-\vec u_2|$, times the scattering
cross section $\sigma$, times the dark matter spatial and velocity
distribution $\rho\big(\vec x, \vec u_2\big)$.  Integrating out
the dark matter velocity, one thus gets for the force acting at
the  point $\vec x(t)$ on the spacecraft trajectory with velocity
$\vec u_1=d\vec x(t)/dt$,
\begin{equation}\label{eq:force}
\delta \vec F= \int d^3 u_2 \langle \delta \vec v_1\rangle |\vec
u_1-\vec u_2| \sigma \rho\big(\vec x, \vec u_2\big)~~~.
\end{equation}
Equating the work per unit spacecraft mass along a trajectory from
$t_i$ to $t_f$ to the change in kinetic energy per unit mass
(assuming that the initial and final times are in the asymptotic
region where the potential energy can be neglected) we get
\begin{align}\label{eq:work}
\delta \frac{1}{2}(\vec v_f^{\,2} -\vec v_i^{\,2}) = &\vec v_f
\cdot \delta \vec v_f = \int_{t_i}^{t_f} dt  (d\vec x/dt) \cdot
\delta \vec F\cr =&\int_{t_i}^{t_f} dt \int d^3 u_2 (d\vec x/dt)
\cdot \langle \delta \vec v_1\rangle |\vec u_1-\vec u_2| \sigma
\rho\big(\vec x, \vec u_2\big)~~~.\cr
\end{align}

\subsection{Cross section and scattering-angle averaged  kinematics}

Let $W$ be the  center of mass scattering energy of the
dark matter-spacecraft nucleon system. A simple calculation shows that to a good
approximation we have
\begin{equation}\label{eq:cmen}
\frac{W}{(m_1+m_2)c^2} \simeq 1+ \frac{m_1 m_2}{2(m_1+m_2)^2} \frac{(\vec u_1-\vec u_2)^2}{c^2}~~~,
\end{equation}
and so for $m_2 \leq m_1$ and for the nonrelativistic velocities $\vec u_1,\, \vec u_2$ of interest,
the scattering is very close to threshold.  Thus the cross section  will be dominated by the lowest partial waves,
which near threshold each have a characteristic power law dependence  on the entrance channel momentum
\begin{equation}\label{eq:kdef}
k=\frac{m_1 m_2}{m_1+m_2} |\vec u_1-\vec u_2|~~~.
\end{equation}

For elastic scattering, the cross section is $S$-wave dominated, and tends to a $k$-independent constant $\sigma_{\rm el}$
near threshold, and the angular average  $2 \langle \sin^2(\theta/2) \rangle$ reduces to
$1 -\langle \cos \theta \rangle =1$.  Thus when Eq. \eqref{eq:el} is substituted into
Eqs. \eqref{eq:force} and \eqref{eq:work}, we can effectively replace $2 \langle \sin^2(\theta/2) \rangle \sigma$
by the $k$-independent constant $\sigma_{\rm el}$.

For exothermic inelastic scattering, the leading contribution
to $\langle \cos \theta \rangle$ comes from the interference term between the $S$- and $P$-waves in the
cross section, which scales  \cite{weinberg} as $k^{-2} k^{1/2} k^{3/2} \sim {\rm constant}$ near threshold.
Writing near threshold
\begin{equation}\label{eq:siginel}
\frac{d\sigma}{d\Omega}= \frac {A_{\rm inel}}{4\pi}  k^{-1} + B_{\rm inel} \frac {3} {4\pi} \cos \theta + ...,
\end{equation}
we have
\begin{align}\label{eq:averages}
\sigma \simeq & A_{\rm inel}  k^{-1}~~~,\cr
\langle \cos \theta \rangle \simeq& B_{\rm inel} /(A_{\rm inel} k^{-1})~~~.\cr
\end{align}
So when Eq. \eqref{eq:inel} for the inelastic exothermic case is substituted into
 Eqs. \eqref{eq:force} and \eqref{eq:work}, we can effectively replace $\langle \cos \theta \rangle \sigma$ by the  $k$-independent
constant $B_{\rm inel}$, remembering, however, that this is not the total cross
section (which approaches  $A_{\rm inel} k^{-1}$ near threshold) but is proportional to the coefficient of the
$S$-wave $P$-wave interference term in the differential cross section.

\subsection{The dark matter distribution function $\rho\big(\vec x, \vec u_2\big)$}

We now address the task of formulating a model for the distribution function $\rho\big(\vec x, \vec u_2\big)$
that describes  dark matter postulated to be in orbit around the earth.  The simplest model would be a disk
composed of dark matter in circular orbits in earth's equatorial plane, but attempts to fit the flyby anomaly data
with such a model were unsuccessful, since for any reasonable disk inner radius, some of the flybys (such as NEAR) pass inside
the disk.    We thus proceed to the next simplest model, which is constructed from
dark matter in a circular orbit, of radius $r$ and tilted at an angle $\psi$ ($0 \leq \psi \leq \pi$) with respect to earth's equatorial
plane.  If the earth were exactly spherically symmetric, its gravitational field would be  strictly monopole,
and such a tilted orbit would be stable.  But in fact the earth's rotation produces an equatorial bulge, and so
its mass distribution is only axially symmetric around its rotation axis, giving rise to quadrupole and higher moments
in its gravitational field.  As a result of these higher moments,  the tilted orbit precesses around
the earth's rotation axis, in such a way that the angular momentum component $L_z$ along the earths's axis is conserved.
Over a long period of time, this precession will smear an initial cluster of tilted orbits into a uniform shell, obtained by
averaging the tilted circle over the azimuthal angle that its normal makes with respect to the earth's rotation axis.

To give this picture a mathematical description, let $x,y,z$ be a Cartesian axis system, with positive $z$ pointing to the
earth's North pole (so that the rotation sense of the earth is from $x$ to $y$).  Let the normal $\hat n$ to the tilted orbit have polar
angle $\psi$ and azimuthal angle $\phi$ with respect to this system, so that $\hat n(\psi,\phi)=(\sin\psi \cos \phi,
\sin\psi \sin \phi, \cos \psi)$,  and let the angle of rotation within the plane
of the dark matter orbit be $\theta$, with increasing $\theta$ corresponding, at $\psi=0$, to the direction of earth's rotation.   Then a parametric description of the tilted circle is $\vec P(r,\theta,\phi)\equiv
\big(P_x(r,\theta,\phi),P_y(r,\theta,\phi),P_z(r,\theta,\phi)\big)$, with
\begin{align}\label{eq:circle_pos}
P_x(r,\theta,\phi)=&\,r(\cos \theta \cos \psi \cos \phi-\sin\theta \sin \phi) ~~~,\cr
P_y(r,\theta,\phi)=&\,r(\cos \theta \cos \psi \sin \phi + \sin \theta \cos \phi)~~~,\cr
P_z(r,\theta,\phi)=&-r \cos \theta \sin \psi~,~~~ |P_z(r,\theta,\phi)|\leq r \sin \psi~~~.\cr
\end{align}
 The corresponding velocity unit vector of a dark matter particle in the tilted circular orbit is
 $\vec U(\theta,\phi)= \big(U_x(\theta,\phi),U_y(\theta,\phi),U_z(\theta,\phi)\big)=r^{-1} d \vec P/d\theta$, with
 \begin{align}\label{eq:circle_vel}
 U_x=&-\sin \theta \cos \psi \cos \phi - \cos \theta \sin \phi~~~,\cr
 U_y=&-\sin \theta \cos \psi \sin \phi + \cos \theta \cos \phi~~~,\cr
 U_z=&~~\sin \theta \sin \psi~~~.
 \end{align}
 The velocity vector is obtained by multiplying the velocity unit vector by the velocity magnitude $ (GM_\oplus/r)^{1/2}$  for  a particle in a circular orbit of radius $r$, with $G$ the Newton gravitational constant and
 $M_\oplus$ the earth mass.

 Integrating the position and velocity distribution for a tilted circular orbit over the angles $\theta,\,\phi$ gives the distribution for the corresponding shell, and integrating over the shell parameters $r,\,\psi $ with a general weighting
function $w(r,\psi)$ gives as the model for the dark matter distribution function
 \begin{equation}\label{eq:shelldens}
 \rho\big(\vec x, \vec u_2\big)=\int dr \int d\psi\, w(r,\psi) \int_0^{2\pi}d\theta \int_0^{2\pi} d\phi \,
 \delta^3\big(\vec x - \vec P(r,\theta,\phi)\big) \delta^3\big(\vec u_2-(GM_\oplus/r)^{1/2} \vec U(\theta,\phi)\big) ~~~,
 \end{equation}
with the corresponding total number of particles in the shell given by
\begin{equation}\label{eq:shellint}
N\equiv \int d^3x \int d^3 u_2 \,\rho\big(\vec x, \vec u_2\big)= 4 \pi^2 \int dr \int d\psi\, w(r,\psi)~~~.
\end{equation}
Referring to Eq. \eqref{eq:work}, we have to evaluate an integral over the distribution function of the form
\begin{equation}\label{eq:orbitint}
I=\int dt \int d^3u_2 F(\vec x(t),d\vec x(t)/dt,\vec u_2) \rho(\vec x(t),\vec u_2)~~~,
\end{equation}
with $F(\vec x(t),d\vec x(t)/dt,\vec u_2)$ given by
\begin{equation}\label{eq:fdef}
F(\vec x(t),d\vec x(t)/dt,\vec u_2) =(d\vec x(t)/dt)\cdot \langle\delta \vec v_1 \rangle|_{\vec u_1=d\vec x(t)/dt}\,\, | d\vec x(t)/dt-\vec u_2| \sigma~~~.
\end{equation}
On substituting Eq. \eqref{eq:shelldens} and noting that the coordinate delta function constrains
$r=|\vec P(r,\theta,\phi)|=|\vec x(t)|\equiv r(t)$,  we obtain
\begin{align}\label{eq:orbitint1}
I=&\int dt \int d\psi \,w(r(t),\psi) \int dr  \cr
\times & \int_0^{2\pi} d\theta \int_0^{2\pi} d\phi\, F\big(\vec x(t), d\vec x(t)/dt, (GM_\oplus/r(t))^{1/2} \vec U(\theta,\phi) \big)  \delta^3\big(\vec x(t) - \vec P(r,\theta,\phi)\big) ~~~.\cr
\end{align}

As shown in Appendix A, by making changes of variable one can carry out the integrations over $r$, $\phi$ and $\theta$ in
Eq. \eqref{eq:orbitint1},
leaving an integral in which
$\theta$ and $z$ have been replaced, by virtue of the delta function constraints, by $\theta(\vec x(t))$ and $z(t)\equiv z(\vec x(t))$,
\begin{align}\label{eq:orbitint3}
I=&\int dt \int d\psi \,w(r(t),\psi)  \sum_{\pm} F\big(\vec x(t), d\vec x(t)/dt, (GM_\oplus/r(t))^{1/2} \vec U_{\pm}(\theta(\vec x(t)),\phi(\vec x(t)))\big) \cr
\times &\frac{1} {r(t)\sqrt{r(t)^2\sin^2\psi - z(t)^2}}~~~.\cr
\end{align}
Note that by virtue of Eq. \eqref{eq:circle_pos}, the integration domain extends only over $|z(t)|\leq r(t) \sin \psi$, and hence the argument of the square root is nonnegative.
In Eq. \eqref{eq:orbitint3}  the sum over $\pm$ is over the two roots $\theta(\vec x(t))$ of the equation $\cos \theta(\vec x(t)) = -z(t)/(r(t) \sin \psi)$,
which differ in the sign of $\sin \theta$,
\begin{equation}\label{eq:roots}
\sin \theta(\vec x(t))  = \pm \sqrt{1-z(t)^2/(r(t)^2 \sin^2\psi)}~~~,
\end{equation}
while the values of $\phi(\vec x(t))$ corresponding to these two roots $\theta(\vec x(t))$  are  obtained by equating $\vec x(t)$ to $\vec P$ and then solving  Eq. \eqref{eq:circle_pos} for $\cos \phi$ and $\sin \phi$.
The two  roots correspond to the fact that a circular orbit with tilt angle $\psi$ consists of two semicircular segments, with opposite directions of the velocity component normal to the equatorial plane.  Thus the  intersection of the  spacecraft trajectory $\vec x(t)$ with the dark matter shell generated by azimuthal rotation of such a tilted circular orbit
will intersect two segments of circular orbits, one up-going and one down-going relative to the equatorial plane.

It will be useful for what follows to express the  unit velocities $\vec U_{\pm}(\theta(\vec x(t)),\phi(\vec x(t))) $ in
terms of their components on unit vectors $\hat n_{\parallel}(t)=(\hat z \times \hat x(t))/|\hat z \times \hat x(t)|$ and $\hat n_{\perp}(t)=\hat x(t) \times \hat n_{\parallel}(t)$, normal to  $\hat x(t)=\vec x(t)/r$, that are respectively parallel (in the sense of earth rotation) and perpendicular to the earth equatorial plane.  A simple calculation given in Appendix B shows that $\vec U_{\pm}$ are given on this basis by
\begin{equation}\label{eq:uonbasis}
\vec U_{\pm}(\theta(\vec x(t)),\phi(\vec x(t))) =C(t) \hat n_{\parallel} \pm D(t) \hat n_{\perp}~~~,
\end{equation}
with the coefficients $C(t)$ and $D(t)$ given by
\begin{equation}\label{eq:cd}
C(t)=\frac {r(t)\cos \psi} {\sqrt{r(t)^2-z(t)^2}}~,~~~
D(t)=\frac{\sqrt{r(t)^2 \sin^2\psi -z(t)^2}} {\sqrt{r(t)^2-z(t)^2}}~~~,
\end{equation}
which obey $C(t)^2 +D(t)^2=1$.  Explicit expressions for $\hat n_{\parallel}(t)$  and $\hat n_{\perp}(t)$ in the flyby
plane basis are given in the next subsection, which together with Eqs. \eqref{eq:uonbasis} and \eqref{eq:cd} give the formulas for the unit velocities
$\vec U_{\pm}(\theta(\vec x(t)),\phi(\vec x(t))) $ on the flyby plane basis needed in the numerical computations.

 \subsection{Flyby orbital plane kinematics}

 The Anderson et al. paper  \cite{anderson} gives the flyby orbit parameters in terms of coordinates on the celestial
 sphere, but it will be more convenient for our purposes to carry out all flyby orbit calculations in
 the flyby orbital plane.  Let $x_o,y_o,z_o$ be a Cartesian axis system, with $z_o$ normal to the flyby orbital
 plane.   The flyby orbit can then be written in parametric form as
 \begin{align}\label{eq:param}
 x_o(t)=&r(t) \cos \theta_o(t)~~~,\cr
 y_o(t)=&r(t) \sin \theta_o(t)~~~,\cr
 r(t)=&\frac {p} {1+e \cos \theta_o(t)},~~~~~~R_f=\frac{p}{1+e}~~~,\cr
 dx_o(t)/dt=& \frac {-V_f \sin \theta_o(t)}{1+e}=\frac{-y_o(t)}{1+e\cos \theta_o(t)} d\theta_o(t)/dt~~~,\cr
 dy_o(t)/dt=&\frac{V_f (e+\cos \theta_o(t))}{1+e}=\frac{er(t)+x_o(t)}{1+e \cos \theta_o(t)} d\theta_o(t)/dt~~~,\cr
 d\theta_o(t)/dt=&\frac{R_fV_f}{r(t)^2}~~~.\cr
 \end{align}
The scale parameter $p$, the eccentricity $e$, the velocity at closest approach to earth $V_f$,
the radius at closest approach $R_f$, and the velocity at infinity $V_{\infty}$ are given in Table I for each of the six flybys discussed in \cite{anderson},
together with the polar angle $I$ and azimuthal angle $\alpha$ of the earth's north pole with respect to the
$x_o,y_o,z_o$ coordinate system.  The quantities $V_f$ and $V_{\infty}$
are given directly in \cite{anderson}, while $R_f$, $p$, and $e$ can be calculated from them using the formulas
\begin{align}\label{eq:ep}
R_f=&\frac{2GM_{\oplus}}{V_f^2-V_{\infty}^2}~~~,\cr
e=&1+\frac{2 V_{\infty}^2}{V_f^2-V_{\infty}^2}~~~,\cr
p=&\frac {4GM_{\oplus}}{V_{\infty}^2} \left[ \left( \frac{V_{\infty}^2}{V_f^2-V_{\infty}^2}\right)^2 +
\frac {V_{\infty}^2}{V_f^2-V_{\infty}^2} \right]~~~.\cr
\end{align}
The earth axis polar angle $I$ is also directly given in \cite{anderson}, while the azimuthal angle $\alpha$
can be calculated from the formula
\begin{equation}\label{eq:alpha}
\cos \alpha=\frac{\sin \phi'}{\sin I}~~~,
\end{equation}
with $\phi'$ the geocentric latitude at closest approach (which is called $\phi$ in \cite{anderson}; with the orbit parametrization of Eq. \eqref{eq:param}, $\phi'$  is the latitude of the positive $x_o$ axis).  This formula does not determine
the quadrant in which $\alpha$ lies, but this can be fixed from the
additional orbital parameters given in \cite{anderson} (with some corrections supplied to me
by J.K. Campbell \cite{campbell}).  Enough orbit parameters are given in \cite{anderson} to provide several redundancies that serve as cross-checks on these calculations.

\begin{table} [b]\label{table:param}
\caption{Flyby orbital parameters}
\centering
\begin{tabular} {|c| c |c| c| c| c |c|}
\hline\hline
~~~&~~~GLL-I~~~ & ~~~GLL-II~~~ & ~~~NEAR~~~ & ~~~Cassini~~~ & ~~~Rosetta~~~ & ~~~Messenger~~~ \\
\hline
$V_f$ (km/s) & 13.740 & 14.080 & 12.739 & 19.026 & 10.517 & 10.389 \\
$R_f$ (km)   &7,334  & 6,674 & 6,911 & 7,544 & 8,332 & 8,715  \\
$V_{\infty}$ (km/s)& 8.949 & 8.877 & 6.851 & 16.010 & 3.863 & 4.056 \\
$e$ & 2.474 & 2.320 & 1.814 & 5.851 & 1.312 & 1.360 \\
$p$ (km) & 25,480 & 22,160 & 19,450 & 51,690 & 19,260 &  20,570 \\
$I$ (deg) & 142.9 & 138.7 & 108.0 & 25.4 & 144.9 & 133.1 \\
$\alpha$ (deg) & -45.1 & -147.4 & -55.1 & -158.4 & -53.1 & 0.0 \\
\hline
\end{tabular}
\end{table}

\vfill\break

To carry out the computation of the flyby velocity change in the flyby plane basis $x_o,y_o,z_o$ we will need the components
of $\hat n_{\parallel}$ and $\hat n_{\perp}$ on this basis.  In Eq. \eqref{eq:otho} we gave their components on the earth centered basis $x,y,z$; these can be rotated to the flyby plane basis, but it is simpler to calculate them directly by going back
to the defining cross product relations, using the components of $\vec x(t)$ and of the
earth axis $\hat z$ on the flyby plane basis,
\begin{align}\label{eq:planebasis}
\vec x(t)=&(x_o(t),y_o(t),0) ~~~,\cr
\hat z=& (\sin I \cos \alpha, \sin I \sin \alpha, \cos I)~~~.\cr
\end{align}
From these we find
\begin{align}\label{eq:otho1}
\hat n_{\parallel}(t)=&\frac{\hat z \times \hat x(t)}{|\hat z \times \hat x(t)     |}=\frac{1}{\sqrt{r(t)^2-z(t)^2}}\big(-y_o(t)\cos I,\,x_o(t)\cos I,\,(y_o(t) \cos \alpha-x_o(t)\sin \alpha)\sin I\big)~~~,\cr
\hat n_{\perp}(t)= &\hat x(t) \times \hat n_{\parallel}(t)=\frac{1}{r(t)\sqrt{r(t)^2-z(t)^2}} \cr
\times &\big(y_o(t)(y_o(t) \cos \alpha - x_o(t) \sin \alpha ) \sin I,\,
-x_o(t)(y_o(t) \cos \alpha - x_o(t) \sin \alpha) \sin I,\,r(t)^2 \cos I\big)~~~,\cr
\end{align}
with
\begin{align}\label{eq:zandr}
r(t)=&|\vec x(t)|=\sqrt{x_o(t)^2 + y_o(t)^2}~~~,\cr
z(t)=&\vec x(t) \cdot \hat z =  (x_o(t) \cos \alpha + y_o(t) \sin \alpha) \sin I~~~.\cr
\end{align}
Substituting Eq. \eqref{eq:param} for $x_o(t)$ and $y_o(t)$ into Eq. \eqref{eq:zandr} we have
\begin{equation}\label{eq:zandr1}
z(t)=r(t)\sin I \cos\big(\theta_o(t)-\alpha\big)~~~,
\end{equation}
which allows one to rewrite the Jacobian factor appearing in Eq. \eqref{eq:orbitint3} as
\begin{equation}\label{eq:jacobian}
\frac{1} {r(t)\sqrt{r(t)^2\sin^2\psi - z(t)^2}}= \frac {1} {r(t)^2 \sin \psi \sqrt{1 - (\sin I / \sin \psi)^2
\cos^2\big(\theta_o(t)-\alpha\big) }}~~~.
\end{equation}
when the argument of the square root is nonnegative.

\subsection{Simplified model used for numerical work}

The model as defined above involves a general weighting function $w(r,\psi)$, but for an initial survey we
make the simplifying assumption of only a single tilt angle $\psi_i$, $\psi_e$  for the inelastic and elastic
scatterers, respectively, and Gaussian distributions in $r$ with different centers and widths  for each.
Thus we take for the inelastic scatterers
\begin{equation}\label{eq:winel}
w_i(r,\psi)=K_i e^{-(r-R_i)^2/D_i^2} \delta(\psi-\psi_i)~~~,
\end{equation}
and for the elastic scatterers
\begin{equation}\label{eq:wel}
 w_e(r,\psi)=K_e e^{-(r-R_e)^2/D_e^2} \delta(\psi-\psi_e)~~~.
\end{equation}
With this choice, the integral of Eq. \eqref{eq:shellint} becomes
\begin{equation}\label{eq:newshellint}
N_{\ell}=4\pi^{5/2} K_{\ell}D_{\ell}~,~~~\ell=i,e~~~.
\end{equation}
 It is now convenient to combine the constants $K_{i,e}$ with the mass-dependent constants
 appearing in Eqs. \eqref{eq:el} and \eqref{eq:inel} of Sec. IIA, and the constants $\sigma_{\rm el}$ and
 $B_{\rm inel}$ introduced in Sec. IIC, giving new parameters $\rho_i,\, \rho_e$ characterizing the
 effective density times cross section for the inelastic and elastic scatterer distributions,
 \begin{align}\label{eq:rhodefs}
  \rho_e\equiv& \frac{m_2}{m_1+m_2} \sigma_{\rm el} K_e~~~,\cr
  \rho_i\equiv&  \Bigg( \frac{2 \Delta m ~m_2'} {
 m_1 (m_1+m_2')}\Bigg)^{1/2} B_{\rm inel} K_i~~~.\cr
 \end{align}
 Thus in Eq. \eqref{eq:orbitint1} we effectively replace (see Eq. \eqref{eq:fdef})
 \begin{equation}\label{eq:replace1}
 \int d\psi \, w(r(t),\psi) \, F(\vec x(t),d\vec x(t)/dt,\vec u_2)
 \end{equation}
 by
 \begin{equation}\label{eq:replace2}
 \sum_{\ell=i,e} \Big\{ |d\vec x(t)/dt-\vec u_2| (d\vec x(t)/dt) \cdot \,\vec V_{\ell}\, \rho_{\ell} \,
 e^{-(r(t)-R_{\ell})^2/D_{\ell}^2} \Big\}|_{\psi=\psi_{\ell}} ~~~,
 \end{equation}
 with $\vec V_{\ell}$ given by
 \begin{align}\label{eq:Vdeff}
 \vec V_i=& \,c\,\big( d\vec x(t)/dt -\vec u_2\,\big)/|d\vec x(t)/dt -\vec u_2|~~~,\cr
 \vec V_e=& - \big(d\vec x(t)/dt -\vec u_2\big)~~~,\cr
 \end{align}
 and with $\vec u_2$ evaluated as $\vec U_{\pm}$ of Eqs. \eqref{eq:orbitint3} and \eqref{eq:uonbasis}.
 The simplified model thus defined has eight parameters, four parameters $\psi_i,\, \rho_i,\, R_i,\,D_i$ characterizing
 the inelastic scatterers, and four parameters $\psi_e,\,\rho_e,\, R_e,\,D_e$ characterizing the elastic scatterers.
 Finally, we note that by combining Eqs. \eqref{eq:newshellint}  and \eqref{eq:rhodefs}, and approximating
 \begin{equation}\label{eq:approxratio}
 \frac{m_2}{m_1+m_2} \sim  \Bigg( \frac{2 \Delta m ~m_2'} {
 m_1 (m_1+m_2')}\Bigg)^{1/2} \sim \frac{m_2}{m_1}~~~,
 \end{equation}
 we find the following estimates for the total mass in the dark matter shells,
 \begin{align}\label{eq:mass}
 M_e\equiv& m_2 N_e= 4 \pi^{5/2} \rho_eD_e m_1/\sigma_{\rm el}~~~,\cr
 M_i\equiv& m_2 N_i= 4 \pi^{5/2} \rho_iD_i m_1/B_{\rm inel}~~~.\cr
 \end{align}

 \section{Numerical Results and Discussion}

 Let us turn now to numerical fitting of the eight parameter model to the flyby anomalies reported in
 \cite{anderson}.  In carrying out the needed integrals over flyby orbits, we replaced the integration
 over $t$ by an integration over orbit angle $\theta_o$, using the expression for $d\theta_o/dt$ given
 in Eq. \eqref{eq:param}.  To utilize integration mesh points efficiently, the integrations were restricted
 to the parts of the orbits where the Gaussian factors $e^{-(r-R_{\ell})^2/D_{\ell}^2}$ were larger than
 $e^{-9}=0.00012$, that is, to the parts of the orbits where  $|r-R_{\ell}|\leq 3 D_{\ell}$.

 In attempting
 to search for good fits with coarse meshes, we found that the infinite jump in the Jacobian factor of
 Eq. \eqref{eq:jacobian} at the dark matter shell edges led to the search program settling on false minima reflecting
 truncation errors, which were unstable with respect to small changes in the integration mesh or fitting
 parameters.  To avoid
 this problem, we replaced the original Jacobian by a smoothed Jacobian, as follows.  Abbreviating
  $W\equiv z(t)^2/(r(t)^2 \sin^2 \psi)$, and using $\Theta$ to denote the usual step function, the original Jacobian contains
  the  function with an infinite jump at $W=1$,
 \begin{equation}\label{eq:original}
 f(W) = \frac {\Theta(1-W)} {\sqrt{1-W}}~~~.
 \end{equation}
 We replaced this by the following function, which is continuous and has a continuous first derivative,
 \begin{align}\label{eq:newjacob}
 f_{\epsilon}(W)=& \frac{1}{\sqrt{1-W}} ~~~~~~{\rm for}~~ W \leq 1-\epsilon~~~~,\cr
 f_{\epsilon}(W)=&\frac{1}{\sqrt{\epsilon}}e^{-P_{\epsilon}(W)} ~~~~~{\rm for}~~ W \geq 1-\epsilon~~~,\cr
 P_{\epsilon}(W)=&-\frac{1}{2\epsilon}(W-1+\epsilon)+\frac{1}{\epsilon^2}(W-1+\epsilon)^2~~~.\cr
 \end{align}
 For our initial searches we took $\epsilon=10^{-2}$.

 Our numerical searches were carried out by minimizing a least squares likelihood function $\chi^2$, defined as
 \begin{equation}\label{eq:chisq}
 \chi^2=\sum_{k=1}^6 (\delta v_{k; {\rm th}}-\delta v_{k; {\rm A}})^2/\sigma_{k; \rm A}^2~~~,
 \end{equation}
 where $k$ indexes the six flybys reported by Anderson et al.  \cite{anderson}, the  $\delta v_{k; {\rm th}}$ are the
 theoretical values of the velocity discrepancies computed from our model, the $\delta v_{k; {\rm A}}$ are
 the observed values for these discrepancies reported in \cite{anderson}, and the $\sigma_{k; \rm A}$ are the corresponding
 estimated errors in these discrepancies given in \cite{anderson}.  Since the quoted $\sigma_{k; {\rm A}}$ values contain
 both systematic and statistical components, a least squares likelihood function is not a true statistical chi square
 function, but having a quadratic form is very convenient for the following reason. Because the theoretical values
 $\delta v_{k; {\rm th}}$ are linear in the dark matter density times cross section parameters $\rho_{i,e}$,
 \begin{equation}\label{eq:linear}
 \delta v_{k; {\rm th}} = \rho_i \delta v_{k; i} + \rho_e \delta v_{k,e}~~~,
 \end{equation}
 with $\delta v_{k; i,e}$ the respective contributions from the inelastic and elastic scatterers computed with
 $\rho_{i,e}=1$,
 the likelihood function is a positive semi-definite quadratic form in these two parameters.  Hence for fixed values of
 the other six parameters $\psi_{i,e},\,R_{i,e},\,D_{i,e}$, the minimization of $\chi^2$ with respect to the
 parameters $\rho_{i,e}$ can be accomplished algebraically by solving a pair of linear equations
 in the  two variables $\rho_{i,e}$,
 with the result
 \begin{align}\label{eq:rhoie}
 \rho_i=&\frac{C_{ee}G_i-C_{ei}G_e}{C_{ii}C_{ee}-C_{ie}C_{ei}}~~~,\cr
 \rho_e=&\frac{C_{ii}G_e-C_{ie}G_i}{C_{ii}C_{ee}-C_{ie}C_{ei}}~~~,\cr
 \end{align}
 with coefficients given  by
 \begin{align}\label{eq:coeffs}
 C_{\ell \,m} = &\sum_{k=1}^6\frac{\delta v_{k; \ell} \delta v_{k; m} }{\sigma_{k; {\rm A}}^2}~,~~~\ell,m=i,e~~~,\cr
 G_{\ell} = &\sum_{k=1}^6 \frac{\delta v_{k; {\rm A}} \delta v_{k; \ell} } {\sigma_{k; {\rm A}}^2}~,~~~\ell=i,e~~~.\cr
 \end{align}
 This has the effect of reducing the parameter space that must be searched numerically from an eight parameter space
 to a six parameter space, which results in a substantial saving of computational effort.

 Our search procedure was then as follows.  Using a very coarse 10 point integration mesh for the model calculation of the flyby velocity changes, and with $\epsilon=10^{-2}$, we surveyed the six
 parameter space in 31 steps of $\pi/32$ for the tilt angles $\psi_{i,e}$, going from $\pi/64$ to $\pi-\pi/64$, in 20 steps of
 2,500 km for the Gaussian centers $R_{i,e}$, going from 15,000 km to 62,500 km, and in 5 steps of 1000 km for the Gaussian
 widths $D_{i,e}$, going from 1,000 km to 5,000 km.  For each of the 9,610,000 steps in this survey, the values of $\rho_{i,e}$ were then
 optimized by using Eqs. \eqref{eq:rhoie} and \eqref{eq:coeffs}, and the resulting data for $\chi^2$ values less than $25$ which also had positive $\rho_e$  were written to a storage file.    This left 18 potential starts for fits. For about a half
 dozen of these, we used the  corresponding  sets of parameter values as
 starting points for a six parameter minimization search using the CERN program Minuit, with successively 200 and
 then 2000 point integration meshes for the
 model calculation of the flyby velocity changes, and using double precision arithmetic throughout (as recommended
 in the Minuit documentation).  Finally, using the optimized parameter values obtained this way, we tested for stability
 of the $\chi^2$ values and resulting fits with respect to program modifications, such as refinement of the integration mesh.  The parameter space survey took several hours on our pentium processor laptop, the
 Minuit minimizations took typically minutes (or less) each, and the stability checks took of the order of seconds.

 This procedure showed that for $\epsilon=10^{-2}$ excellent fits could be obtained with a wide range of values of the radius $R_i$ of the inelastic dark matter scatterer shell.  Using the parameters for these good fits as a starting point, we then did a series of 5 parameter fits, each for a different fixed value of $R_i$.  Also using the good fits as starting points, we did a similar series of
5 parameter fits, versus fixed $R_i$, this time with $\epsilon=10^{-16}$ corresponding to no smoothing of the Jacobian
discontinuity (up to the accuracy of
double precision truncation errors), but using an adaptive integration program to adequately sample points on the trajectories where the Jacobian
becomes large.  These searches (as well as a 6 parameter fit in the $\epsilon=10^{-16}$ case) show that the model with no
smoothing has a distinct $\chi^2$ minimum at $R_i=34,520$ km.  Results in both $\epsilon$ cases are given in Tables II -- IV.
We caution that the $\epsilon=10^{-2}$ cases do not exactly obey the constraints between dark matter position and velocity
required by orbital dynamics, so it is not clear at this point whether the wide range of $R_i$ values and nearly exact fits
obtained in this case are a reflection of just the smoothing, which will be present in a more realistic dark matter orbit model,
or are an artifact associated with relaxing the orbital constraints.

 From the products $\rho_iD_i$ and $\rho_eD_e$ for each fit, one can use Eq. \eqref{eq:mass} to estimate the total mass in
 the dark matter shells, in terms of the elastic and inelastic scattering parameters $\sigma_{\rm el}$ and $B_{\rm inel}$.
 Alternatively, given the upper bound \cite{adler2} on the mass of dark matter in orbit around the earth between the LAGEOS satellite orbit
 and the moon's orbit, of $4 \times 10^{-9} M_{\oplus} \sim 1.4 \times 10^{43} {\rm GeV}/c^2$, one can turn these relations into
 lower bounds on $\sigma_{\rm el}$ and $B_{\rm inel}$.  For example, from  the values $\rho_iD_i=0.00304\,{\rm km}^2$ and $\rho_eD_e=19.2\, {\rm km}^2$ for fit 2d, one finds the bounds
\begin{align}\label{eq:bounds}
\sigma_{\rm el} \geq& 9.4\times 10^{-31} {\rm cm}^2~~~,\cr
B_{\rm inel} \geq& 1.5 \times 10^{-34} {\rm cm}^2~~~,\cr
\end{align}
which are consistent with the cross section range arrived at from various constraints in \cite{adler1}.  The spatial constraints
found in \cite{adler1}, which require that the dark matter should be localized well away from the earth and the moon, are also obeyed.

In Table V we give the results of fitting the data with $R_i$
constrained to the value 34,520 km found in fit 2d (repeated in the
first line of this table), versus increasing values of the Gaussian
width $D_i$.  These results, together with those for fits 2e--g in
Tables II and IV, show that the range of widths $D_i$ for good fits
extends up to around 10,000 km. The fact that $D_i$ is not
well-determined is also seen in the calculation leading to  Table
VII, where in fit 4a we give the result of repeating fit 2d with a
refined (4000 point) integration mesh. The parameter values for fit
4a agree to within 1\% with those of fit 2d, except for $D_i$, which
in fit 4a is 2030 km, and $10^6 \times \rho_i$, which in fit 4a is
1.49 km, with the product $\rho_i D_i$ matching that of fit 2d to
within 1\%.

 In Table VI, we show the results of basing the fit solely on a
shell of inelastic scatterers, without a second shell of elastic
scatterers. As seen, with this restriction it is not possible to get
good fits, even when various combinations of the flyby data are
excluded from the fits. For example, as shown on the last line of
Table VI, the four parameter model with only inelastic scatterers
cannot give a good fit to just the two flyby data points from NEAR
and Messenger. In Table VII, following up on a suggestion by V. Toth
\cite{toth}, we give the results of fitting the full model, with
both elastic and inelastic scatterers,  to the flyby data, with one
flyby at a time omitted from the fit.  These results show that the
predicted value for the anomaly of each omitted flyby is in qualitative
accord with the experimental value.

 The results in Tables II -- VII show that the dark matter scattering model, with inelastic and elastic scatterers,  can account for the flyby anomaly data.  One could
 argue that the fits are too good, and are indicative of ``over-fitting'', since there are 8 parameters in the model (9 if one
 includes $\epsilon$ in the smoothed case), and only
 6 data points.  On the other hand, it was not a priori obvious that such a simple model should be able to account for data from
 a complicated physical process with a three-dimensional geometry, and the results shown in Table VII support the view that the
 success of the model is not attributable to over-fitting of the data.    Further steps in this investigation would be: (1)
 incorporation of further flybys into the fits, when the flyby parameters in Table I and the corresponding velocity
 discrepancy and error values are available, or alternatively, using fit 2d (or 4a) to predict  the velocity discrepancy for
 future flybys, given their orbital parameters; (2) incorporating constraints on residual drag coming from fitting satellite drag measurements to conventional drag sources; (3) as suggested to me by V. Toth \cite{toth}, incorporating the time development of the velocity anomaly near perigree when such
data becomes available from improved tracking of future flybys; (4) as suggested to me by J. Rosner \cite {rosner}, investigating possible constraints arising from the effect of the quadrupole moment of the dark matter shells on the precession of high-lying satellite orbits;  (5) extending the model to include a general form of the weighting function $w(r,\psi)$; and (6) extending the model to include shells generated by precessing elliptical, as opposed to circular orbits, and shells generated by a precessing Schwarzschild disk \cite{tremaine}.  The  extensions (5) and (6), which can incorporate consistent smoothing of the Jacobian,  will require computing resources well beyond those used here to analyze the 8 parameter model.  It will also be necessary to
address the question of mechanisms for producing dark matter shells.  According to A. Peter \cite{apeter}, the accumulation cascade suggested in \cite{adler1} is not viable as a mechanism.  Another scenario, suggested by Dr. Peter's comments and the structure of
the  model formulated here, would involve the gravitational capture by the earth of a dense (up to $ \sim 10^{15}$ times galactic halo mean density, that is $\sim 10^{-9}$ times
mean ordinary matter density) condensed ball of dark matter into an orbit tilted with respect to earth's rotation axis; breakup of this by tidal forces could then
lead to population of a shell of the type we have assumed.\footnote{The constraints derived in \cite{adler1} on the sun-bound
dark matter density are not relevant for this scenario for producing dark matter shells.}  If the flyby anomalies are ultimately confirmed, detailed study
of such a mechanism would be warranted.

\section{Acknowledgements}
This work was
supported by the Department of Energy under grant no
DE-FG02-90ER40542, and parts of this work were done during the author's stay at
the Aspen Center for Physics.   I wish to thank Scott Tremaine for helpful conversations about orbital
dynamics,  James Campbell for sending me corrections to some of the data published in \cite{anderson},
Michele Papucci for suggesting that I use the CERN minimization program Minuit, and Prentice Bisbal
for assistance in downloading it to my computer.  I also wish to thank Angelo Bassi, Annika Peter, Jonathan Rosner, and Viktor Toth  for helpful comments after the initial version of this paper was posted on the arXiv.

\appendix
\section{Changes of variable to integrate out the spatial delta function}

To eliminate the spatial delta function in Eq. \eqref{eq:orbitint1}, we note that rewriting $\vec P$ in terms of spherical coordinates,
\begin{equation}\label{eq:sphercoord}
\vec P=r(\sin \omega \cos \beta, \sin \omega \sin \beta, \cos \omega)~~~,
\end{equation}
the delta function $\delta^3(\vec x-\vec P)$ becomes
\begin{equation}\label{eq:spherdelt}
\delta^3(\vec x -\vec P)= r^{-2}|\sin \omega|^{-1}  \delta(|\vec x|-r) \delta(\omega(\vec x)-\omega)
\delta(\beta(\vec x) -\beta)~~~.
\end{equation}
Equating Eq. \eqref{eq:sphercoord} with Eq. \eqref{eq:circle_pos}, we see that
\begin{align}\label{eq:equivs1}
\beta=&\phi+\Psi(\theta,\psi)~~~,\cr
\cos \Psi(\theta,\psi) = &\frac {\cos \theta \cos \psi}{\sqrt{1-\cos^2\theta \sin^2 \psi}}~,~~~
\sin \Psi(\theta,\psi) = \frac{\cos \theta} {\sqrt{1-\cos^2\theta \sin^2 \psi}}~,~~~\cr
\end{align}
and
\begin{equation}\label{eq:equivs2}
\cos \omega = -\cos \theta \sin \psi~~~.
\end{equation}
Using Eq. \eqref{eq:equivs1},  on substituting Eq. \eqref{eq:spherdelt}
with $\vec x=\vec x(t)$
into Eq. \eqref{eq:orbitint1}, we can immediately eliminate the $r$ and $\phi$ integrations, leaving
\begin{align}\label{eq:orbitint2}
I=&\int dt \int d\psi \,w(r(t),\psi) r(t)^{-2}
 \int_0^{2\pi} d\theta |\sin \omega|^{-1}\cr
 \times & F\big(\vec x(t), d\vec x(t)/dt, (GM_\oplus/r(t))^{1/2}\vec U(\theta,\phi(\vec x(t))) \big)  \delta(\omega(\vec x)-
 \omega) ~~~.\cr
\end{align}

To carry out the $\theta$ integration, we differentiate Eq. \eqref{eq:equivs2}, giving
\begin{align}\label{eq:Jacobian}
\frac{d\theta}{\sin \omega}=&\frac {-d\omega}{\sin \psi \sin \theta} \cr
=&\frac{-d\omega}{\sqrt{\sin^2 \psi (1-\cos^2 \theta)}} = \frac{-d\omega}{\sqrt{\sin^2 \psi-\cos^2 \omega}}  \cr
=& \frac{-rd\omega} {\sqrt{r^2\sin^2\psi - z^2}}~~~,\cr
\end{align}
with $z=r \cos \omega$.  Substituting this into Eq. \eqref{eq:orbitint2}, we can carry out the $\theta$ integral, leaving an the integral given in Eq. \eqref{eq:orbitint3} of the text.

\section{Calculation of the coefficients $C(t)$ and $D(t)$}

From the defining cross product relations, we see that on the geocentric basis system with $z$ aligned along the earth rotation axis, we have
\begin{align}\label{eq:otho}
\hat x(t)=&\frac{1}{r(t)}(x(t),y(t),z(t)) ~~~,\cr
\hat n_{\parallel}(t)=&\frac{\hat z \times \hat x(t)}{|\hat z \times \hat x(t)     |}=\frac{1}{\sqrt{r(t)^2-z(t)^2}}(-y(t),x(t),0)~~~,\cr
\hat n_{\perp}(t)= &\hat x(t) \times \hat n_{\parallel}(t)=\frac{1}{r(t)\sqrt{r(t)^2-z(t)^2}}
\big(-x(t)z(t),-y(t)z(t),r(t)^2-z(t)^2\big)~~~.\cr
\end{align}
To express the unit velocity of Eq. \eqref{eq:circle_vel} on this basis, at the intersections where $\vec x(t) =
\vec P(r,\theta,\phi)$, we rewrite Eq. \eqref{eq:circle_pos} as
\begin{align}\label{eq:circle_pos1}
x(t)/r(t)=&\,(\cos \theta \cos \psi \cos \phi-\sin\theta \sin \phi) ~~~,\cr
y(t)/r(t)=&\,(\cos \theta \cos \psi \sin \phi + \sin \theta \cos \phi)~~~,\cr
z(t)/r(t)=&- \cos \theta \sin \psi~~~.\cr
\end{align}
The third of these equations determines $\cos \theta$ and $\sin\theta$ in terms of $\vec x(t)$,
\begin{align}\label{eq:cos_sinthet}
\cos\theta= &-z(t)/(r(t) \sin \psi)~~~,\cr
\sin\theta = &\pm \sqrt{1-z(t)^2/(r(t)^2 \sin^2\psi}~~~,\cr
\end{align}
while solving the first two gives $\sin \phi$ and $\cos \phi$ in terms of $x(t)$ and $y(t)$,
\begin{align}\label{eq:solving}
\sin \phi= &\frac{gy(t)-hx(t)}{r(g^2+h^2)}~~~,\cr
\cos \phi= &\frac{gx(t)+hy(t)}{r(g^2+h^2)}~~~,\cr
\end{align}
with $g=\cos \theta \cos \psi$,  $h=\sin \theta$, which obey
\begin{equation}\label{eq:ghsquare}
g^2+h^2=1-\cos^2 \theta \sin^2 \psi = 1-z(t)^2/r(t)^2~~~.
\end{equation}
Substituting Eqs. \eqref{eq:cos_sinthet} and \eqref{eq:solving} into \eqref{eq:circle_vel} gives the
velocity components at the intersections expressed in terms of $\vec x(t)$, and comparing with Eq. \eqref{eq:otho}
then  identifies the coefficients $C(t)$ and $D(t)$ appearing in the decomposition of $\vec U_{\pm}$ on the
intrinsically defined basis $\hat n_{\parallel}$ and $\hat n_{\perp}$.

\vfill\break

\begin{table} [t]\label{table:fits}
\caption{Flyby anomaly fits}
\centering
\begin{tabular} {|c|c| c |c| c| c| c |c|}
\hline\hline
~~~&~~~ $\chi^2$~~~&~~~GLL-I~~~ & ~~~GLL-II~~~ & ~~~NEAR~~~ & ~~~Cassini~~~ & ~~~Rosetta~~~ & ~~~Messenger~~~ \\
\hline
$\delta v_{\rm A}$ (mm/s) & & 3.92 & -4.6 & 13.46 & -2 & 1.80 & 0.02 \\
$\sigma_{\rm A}$     (mm/s)&  &0.3  & 1.0 & 0.01 & 1 & 0.03 & 0.01 \\
\hline
$\delta v_{\rm th}$~ fits 1a--e  &$<10^{-6}$  & 3.92 & -4.60 & 13.46 & -2.00 & 1.80 & 0.020 \\
$\delta v_{\rm th}$ ~  fit 2a & 2.07   & 3.98 & -5.5 & 13.46 & -3.1 & 1.79 & 0.021 \\
$\delta v_{\rm th}$ ~ fit 2b & 1.68& 4.15     & -5.2& 13.46 & -2.9 & 1.80 &  0.020 \\
$\delta v_{\rm th}$ ~ fit 2c &1.29  & 4.13    & -5.0 & 13.46 & -2.8 & 1.80 & 0.020 \\
$\delta v_{\rm th}$ ~ fit 2d  &0.51 & 3.90    & -4.6& 13.46 & -2.7 & 1.80& 0.020\\
$\delta v_{\rm th}$ ~ fit 2e &0.52  & 3.88    & -4.6& 13.46 & -2.7 & 1.80& 0.020\\
$\delta v_{\rm th}$ ~ fit 2f &0.70  & 3.84    & -4.7& 13.46 & -2.7 & 1.80& 0.021\\
$\delta v_{\rm th}$ ~ fit 2g &7.5  & 3.76     & -4.7& 13.46 & -2.8 & 1.73& 0.028\\
\hline
\end{tabular}

\bigskip

Fits 1a--e are for the smoothed model with $\epsilon=10^{-2}$ and trapezoidal integration,
resulting from a five-parameter fit with $R_i$ constrained to the values shown in Table III.
Fits 2a--g are for the un-smoothed model ($\epsilon=10^{-16}$, which is below truncation errors) and adaptive trapezoidal
integration, resulting from a
\leftline{five-parameter fit with $R_i$ constrained to the values shown in Table IV.}

\end{table}

\begin{table} []\label{table:param1}
\caption{Parameter values for fits 1a--e}
\centering
\begin{tabular} {|c|c| c |c| c| c| c |c|c|}
\hline\hline
~fit~&$10^6 \times \rho_i$~(${\rm km}$)~~& $10^2 \times \rho_e$~(${\rm km}$)~~&$\psi_i$~(rad)~~&$\psi_e$~(rad)~~&$R_i$~(${\rm km}$)~~&$D_i$~(${\rm km}$)~~&$R_e$~(${\rm km}$)~~&$D_e$~(${\rm km}$)~~\\
\hline
1a&0.304  & 0.268  & 1.926  & 0.3939  & 30000  & 6278 & 28620  & 6303   \\
1b&1.55   & 0.245  & 1.261  & 0.3945  & 40000  & 2185 & 27985  & 5890   \\
1c&0.411  & 0.261  & 1.374 & 0.3952  & 50000  & 13540 & 28450  & 6299   \\
1d&0.351  & 0.253  & 1.381  & 0.3946  & 60000  & 20193 & 28340  & 6334   \\
1e&0.343  & 0.248  & 1.394  & 0.3942  & 70000  & 25780 & 28240  & 6367   \\

\hline
\end{tabular}

\bigskip

\end{table}

\begin{table} []\label{table:param2}
\caption{Parameter values for fits 2a--g}
\centering
\begin{tabular} {|c|c| c |c| c| c| c |c|c|}
\hline\hline
~fit~&$10^6 \times\rho_i$~(${\rm km}$)~~& $10^2\times\rho_e$~(${\rm km}$)~~&$\psi_i$~(rad)~~&$\psi_e$~(rad)~~&$R_i$~(${\rm km}$)~~&$D_i$~(${\rm km}$)~~&$R_e$~(${\rm km}$)~~&$D_e$~(${\rm km}$)~~\\
\hline
2a& 0.537  & 0.323  & 1.767  & 0.3902  & 25000  & 3030 & 29370  & 6678   \\
2b& 0.827  & 0.316  & 1.626  & 0.3902  & 30000  & 3030  & 29370  & 6678    \\
2c& 0.965  & 0.309  & 1.515  & 0.3902  & 32500  & 3030 & 29370  & 6678    \\
2d& 1.000  & 0.288  & 1.372  & 0.3902  & 34520  & 3030 & 29370  & 6678    \\
2e& 0.655  & 0.288  & 1.369  & 0.3902  & 35000  & 4663 & 29370  & 6678    \\
2f& 0.348  & 0.288  & 1.364  & 0.3902  & 37500  & 9223 & 29370  & 6678    \\
2g& 0.290  & 0.286  & 1.361  & 0.3902  & 40000  & 11681 & 29370  & 6678    \\
\hline
\end{tabular}

\bigskip

\end{table}

\begin{table} [t]\label{table:fits1}
\caption{Flyby anomaly fits with $R_i=34520$ and indicated values of $D_i$ }
\centering
\begin{tabular} {|c|c| c |c| c| c| c |c|}
\hline\hline
~~~&~~~ $\chi^2$~~~&~~~GLL-I~~~ & ~~~GLL-II~~~ & ~~~NEAR~~~ & ~~~Cassini~~~ & ~~~Rosetta~~~ & ~~~Messenger~~~ \\
\hline
$\delta v_{\rm A}$ (mm/s) & & 3.92 & -4.6 & 13.46 & -2 & 1.80 & 0.02 \\

\hline

$\delta v_{\rm th}$ ~ $D_i=3030$  &0.51 & 3.90   & -4.8& 13.46 & -2.7 & 1.80& 0.02\\
$\delta v_{\rm th}$ ~ $D_i=6060$ &0.68  & 3.94    & -4.8& 13.46 & -2.8 & 1.80& 0.02\\
$\delta v_{\rm th}$ ~ $D_i=9090$ &1.3  & 3.97    & -5.1& 13.46 & -3.0 & 1.80& 0.02\\
$\delta v_{\rm th}$ ~ $D_i=12120$ &4.2  & 3.88    & -4.7& 13.46 & -4.0 & 1.80& 0.02\\
\hline
\end{tabular}

\bigskip

\end{table}

\begin{table} [t]\label{table:fits2}
\caption{Flyby anomaly attempted fits with only inelastic scatterers}
\centering
\begin{tabular} {|c|c| c |c| c| c| c |c|}
\hline\hline
~~~&~~~ $\chi^2$~~~&~~~GLL-I~~~ & ~~~GLL-II~~~ & ~~~NEAR~~~ & ~~~Cassini~~~ & ~~~Rosetta~~~ & ~~~Messenger~~~ \\
\hline
$\delta v_{\rm A}$ (mm/s) & & 3.92 & -4.6 & 13.46 & -2 & 1.80 & 0.02 \\

\hline
$\delta v_{\rm th}$ ~  fit 3a & $0.63\times 10^5$   & 1.87 & 1.8 & 13.0 & 1.5 & 2.9 & 2.4 \\
$\delta v_{\rm th}$ ~ fit 3b & $0.63 \times 10^5$& 1.87     &--& 13.0 & -- & 2.9 &  2.4 \\
$\delta v_{\rm th}$ ~ fit 3c &$0.16 \times 10^4$  & 1.93    & -- & 13.4 & -- & 3.0 & -- \\
$\delta v_{\rm th}$ ~ fit 3d  &$0.61 \times 10^5$ & --    & --& 13.0 & -- & --& 2.5\\

\hline
\end{tabular}

\bigskip

Fit attempts with only inelastic scattering; entries labeled -- were
excluded from the corresponding fit. \leftline{ To two decimal
places, all fits in this table correspond to the parameter values
$\rho_i=0.14$, $\psi_i=1.13$,} \leftline{$R_i=40000$, and
$D_i=2000$.}

\end{table}

\begin{table} [t]\label{table:fits3}
\caption{Flyby anomaly  fits to five of the six flybys}
\centering
\begin{tabular} {|c|c| c |c| c| c| c |c|}
\hline\hline
~~~&~~~ $\chi^2$~~~&~~~GLL-I~~~ & ~~~GLL-II~~~ & ~~~NEAR~~~ & ~~~Cassini~~~ & ~~~Rosetta~~~ & ~~~Messenger~~~ \\
\hline
$\delta v_{\rm A}$ (mm/s) & & 3.92 & -4.6 & 13.46 & -2 & 1.80 & 0.02 \\

\hline
$\delta v_{\rm th}$ ~  fit 4a&0.49 & 3.90  & -4.6 & 13.46 & -2.7& 1.80 & 0.02  \\
$\delta v_{\rm th}$ ~ fit 4b &0.45& {\bf 3.71}    &-4.4& 13.46 & -2.6 & 1.80 & 0.02\\
$\delta v_{\rm th}$ ~ fit 4c &0.49 & 3.91    & {\bf -4.6} & 13.46 & -2.7 & 1.80 & 0.02 \\
$\delta v_{\rm th}$ ~ fit 4d  &0.40 & 3.93    & -4.4&{\bf 16.03} & -2.6 & 1.80& 0.02\\
$\delta v_{\rm th}$ ~  fit 4e & $0.63\times 10^{-3}$   & 3.92 & -4.6 & 13.46 & {\bf -2.7} & 1.80 &0.02\\
$\delta v_{\rm th}$ ~ fit 4f & $0.40\times 10^{-1}$& 3.93     &-4.5& 13.46 & -2.2 &{\bf 1.62} & 0.02 \\
$\delta v_{\rm th}$ ~ fit 4g &0.19  & 3.94    & --4.3 & 13.46 & -2.3 & 1.80 & {\bf 0.12} \\

\hline
\end{tabular}

\bigskip

Fit 4a is a fit with all six flybys included. Fits 4b--4g are fits
with one flyby at a time excluded; the {\it predicted} value for the
flyby omitted in each fit is in boldface. These fits use a factor of
2 finer mesh than \leftline{fit 2d.}

\end{table}


\begin{thebibliography}    {99}
\bibitem{adler1} S. L. Adler,  Phys. Rev. D {\bf 79}, 023505 (2009).

\bibitem{anderson}  J. D. Anderson, J. K. Campbell, J. E. Ekelund, J. Ellis,
and J. F. Jordan, Phys. Rev. Lett. {\bf 100}, 091102 (2008).

\bibitem{weinberg}  S. Weinberg, ``The Quantum Theory of Fields, Vol. I Foundations'', Cambridge University Press (1995), pp.
156-157.

\bibitem{campbell} J. K. Campbell, private email communication (2008).

\bibitem{adler2}  S. L. Adler, J. Phys. A: Math. Theor. {\bf 41}, 412002 (2008).

\bibitem{toth}  V. Toth, private email communication (2009).

\bibitem{rosner} J. Rosner, private email communication (2009).

\bibitem{tremaine}  J. Binney and S. Tremaine, {\it Galactic Dynamics}, second edition,  Princeton University Press (2009),
Sec. 4.4.3.

\bibitem{apeter} A. Peter, private email communication (2009).



\end{thebibliography}
 \end{document}